\newcommand{\cOne}{C} % Matt wants another symbol here, maybe c_1
\newcommand{\addParIfLineNumbers}{\ifLineNumbers{\par\vskip-\baselineskip}\fi}

\documentclass[final,5p,times,sort&compress,twocolumn]{elsarticle} 

\usepackage{graphicx}
\usepackage{amsmath,amssymb}
\usepackage{ifthen}
\usepackage{lineno}
\ifthenelse{\jtype=5}{\nolinenumbers}{\linenumbers}

\journal{Nuclear Instruments and Methods A}

\begin{document}

\begin{frontmatter}

\title{A new method to reduce the statistical and systematic uncertainty of
chance coincidence backgrounds measured with waveform digitizers}

\author{J.~M.~O'Donnell}
\address{Los Alamos National Laboratory, Los Alamos, NM 87544, USA
\ead{odonnell@lanl.gov}}

\begin{abstract}

A new method for measuring chance-coincidence backgrounds during the
collection of coincidence data is presented.  The method relies on
acquiring data with near-zero dead time, which is now realistic due
to the increasing deployment of flash electronic-digitizer (waveform
digitizer) techniques.  An experiment designed to use this new method
is capable of acquiring more coincidence data, and a much reduced
statistical fluctuation of the measured background.  A statistical
analysis is presented, and used to derive a figure of merit for the new
method.  Factors of four improvement over other analyses are realistic.
The technique is illustrated with preliminary data taken as part of a
program to make new measurements of the prompt fission neutron spectra
at Los Alamos Neutron Science Center.  It is expected that the these
measurements will occur in a regime where the maximum figure of merit
will be exploited.

\end{abstract}

\begin{keyword}
chance coincidence background
\sep
waveform
digitizer
\sep
statistical uncertainty
\sep
systematic uncertainty
\end{keyword}

\end{frontmatter}

\section{Introduction and Motivation}

The impact of backgrounds on the statistical significance of an experiment
is of such importance that Knoll discusses it in his seminal book,
{\it Radiation Detection and Measurement}, just as soon as sufficient
statistical background is presented to understand the subject\cite{knoll}.
One of the compromises in performing an experiment is to reduce the amount
of time spent taking foreground data (and losing counts and statistics),
to take background data.  If too much time is spent on the background
measurement, then valuable foreground statistics are lost.  On the
other hand, if insufficient background data are obtained, the quality
of the foreground data is compromised by a poor background subtraction.
In the book, Knoll presents an expression for the optimal division of time
between the two stages of data collection.  The question of systematic
changes between the two separate measurements is not considered~--- but
is often apparent, for example if a background normalization factor has
to be applied during the data analysis.  The complications introduced
by backgrounds warrant all reasonable attempts to remove the background.

Over the years the coincidence technique has proven to be a powerful
method to reduce or even remove large backgrounds.  Knoll also discusses
the nature of chance coincidences as a source of background in coincidence
experiments, and presents a simple formula to estimate the chance
coincident rate.  The formula is traditionally used to help design
layouts for experiments, and provide estimates for the quality of the
data when setting up an experiment.  The formula for the rate of chance
coincidences, $r_\gamma$, between two detectors, counting at rates $r_a$
and $r_b$, during a coincidence time window with width, $\Delta_t$, is
\begin{equation}\label{rcCounts}
r_\gamma = r_a r_b \Delta_t.
\end{equation}

A key concept in the derivation of Eq.~\eqref{rcCounts} is the dead time,
$t_d$, of the detector, electronics and the data acquisition system
({\sc DAQ}), even though $t_d$ is not explicit in the formula.
Eq.~\eqref{rcCounts} is only valid if $r_a t_d \ll 1$ and $r_b t_d \ll 1$.

While Eq.~\eqref{rcCounts} is well known, we are not aware of any analyses
using it to measure the detailed background shape in a coincidence
measurement, presumably due to the difficulty in measuring $r_a$ and
$r_b$ reliably.  High singles rates impose challenges in acquiring,
saving and analyzing complete data sets using an event-triggered {\sc DAQ}
with very real dead-time concerns.

The current work is motivated by the increasing use of flash
digitizers running semi-continuously (waveform digitizers),
with waveform analysis capabilities on board, to build high
throughput {\sc DAQ}s with no contribution to dead time (see
e.g.~\cite{caenAN2503,caenAN2506,caenAN2508}).  When combined with
computers with large storage areas it is now realistic to record
complete data sets of all the signals from all the detectors with dead
time due only to signal overlap (pileup) in the detector and preamps.
Coincidences are identified, in software, later in the analysis.

With such a {\sc DAQ} we show that Knoll's expression to estimate chance
coincidence rates can now be used to determine a bin-by-bin measurement of
the background obtained simultaneously with the foreground.  In addition
to reducing systematic uncertainties, this allows to obtain the maximal
signal statistics, and it will be shown yields a very small statistical
uncertainty on the background measurement.  The question of how much
time to spend on the foreground and background measurements becomes
trivial~--- do both all the time!

Before proceeding, we rewrite Eq.~\eqref{rcCounts} in terms of the counts
observed in an experiment, which also allows us to easily quantify the
statistical uncertainty.  The numbers of counts, $a$ and $b$, obtained in
each detector, and summed over $N$ measurements, are then related to the
measured singles rates by $a = r_a N \Delta_t$ and $b = r_b N \Delta_t$,
and the number of chance coincidences, $\gamma$, in the time-difference
window, $\Delta_t$, is then
\begin{equation}\label{cCounts}
\gamma \pm \sigma_\gamma = {\frac{a b}{N}}
\pm \sqrt{\frac {a b(a + b)} {N^2}},
\end{equation}
where $\sigma_\gamma$ is the statistical uncertainty on $\gamma$,
assuming Gaussian statistics for $a$ and $b$.

In the next two sections we briefly describe an experiment being
developed to measure the prompt fission neutron spectrum ({\sc PFNS})
on $^{239}$Pu\cite{chinuPaper}, and then in more detail, two possible
analysis procedures which use Eq.~\eqref{cCounts} to estimate backgrounds
for this experiment.  In the following section we contrast this approach
with more traditional methods for measuring backgrounds~--- resulting
in the derivation of a figure of merit for the new method of analysis.
There then follows a comment on the data-acquisition requirements to
obtain a data set which is complete enough to apply the current technique
(and a numerical validation of $r_a t_d \ll 1$ and $r_b t_d \ll 1$ for the
example experiment).  Further formulas derived from Eq.~\eqref{cCounts},
and which are needed to implement a full analysis, are presented
in a short series of appendices.

\section{Example Experiment With Complicated Backgrounds}

Data for outgoing fission neutrons from neutron induced fission of
$^{239}$Pu are used to illustrate the techniques discussed here.
The coincidence nature of the experiment arises from detecting one of
the fission fragments together with a neutron generated in the fission
process.  The detection of a fission fragment implicates the timing
properties of the incoming neutron.

\begin{figure}[t]
\includegraphics[angle=-90,width=\columnwidth]{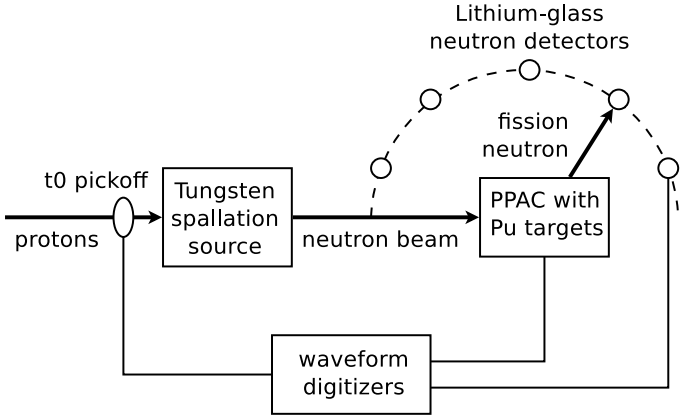}
\caption{ \label{figSetup}
Conceptual layout for a two-arm time of flight $^{239}$Pu$(n;f,xn)$
experiment, detecting fission fragments in a parallel-plate avalanche
counter({\sc PPAC}), (which implicates the timing of an incoming
neutron, with respect to the neutron production time signal from the
t$_0$ pick off), and fission neutrons are detected in the lithium glass
neutron detectors.}
\end{figure}
\begin{figure}[t]
\includegraphics[angle=-90,width=\columnwidth]{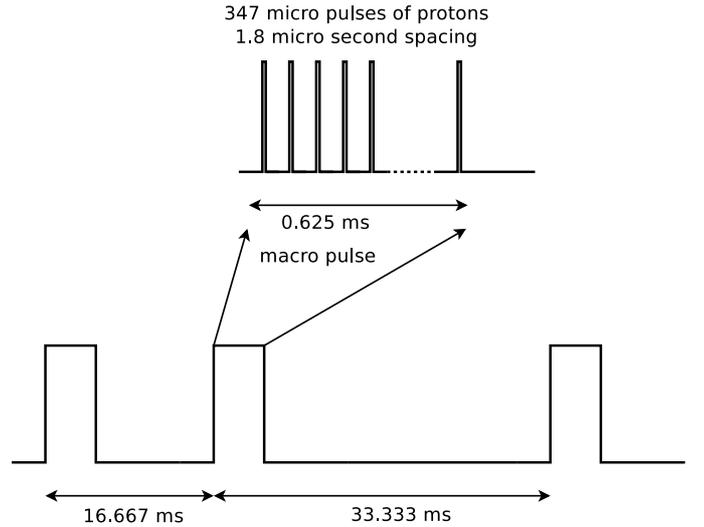}
\caption{ \label{figBeamStructure}
Structure of the proton beam to {\sc WNR}.}
\end{figure}
The data were acquired as part of the development for the {\sc PFNS}
measurements currently being performed at the {\sc WNR/LANSCE} neutron
spallation source\cite{lanscePaper}. A conceptual figure of the experiment
is shown in fig.~\ref{figSetup}.  Beam delivery was structured on two time
scales (fig.~\ref{figBeamStructure}): micro pulses of protons (each making
their own timing signal, $t_0$, and forming their source of spallation
neutrons) occur at nominal 1.8$\mu$s intervals; and macro pulses,
being groups of 347~micro pulses, separated by several milliseconds.
Only two of every three macro pulses were delivered to {\sc WNR}.

Model {\sc ZT4441-DP-PXI} digitizers\cite{zt4441Paper} (sample rate
400Ms$^{-1}$) were used to acquire 1ms long waveforms spanning the
duration of complete macro pulses.  Triggers for most of the digitizers
were distributed across a {\sc PXI} bus from a self-triggering digitizer
receiving the $t_0$ signal.  {\sc ZT1000PXI} cards\cite{zt4441Paper}
enabled distribution of the trigger and a stable reference clock between
{\sc PXI} crates.  Up to 22 lithium-glass detectors and digitizer channels
were used to detect the outgoing neutrons\cite{CfLiglPaper}.  A further
ten digitizer channels processed the signals from a multi-segment
parallel-plate avalanche counter ({\sc PPAC}) for detecting fission
fragments\cite{PPACref}.  Each waveform was analyzed on board the
digitizer using custom firmware, to generate a list of parameters such
as a time stamp, the baseline height, and two integrals of the peak area
at different time offsets from the peak position, for all the peaks
found in the waveform.  The lists of parameters were read from each
digitizer across the {\sc PXI} bus into computer memory for storage and
further processing, before preparing the digitizers to process another
macro pulse.

The times within a macro pulse of individual peaks in the $t_0$, {\sc
PPAC} and lithium-glass signals are denoted as $t_0$, $t\!_f$, and $t_n$,
respectively; and the lists of these times as reported by the digitizers
are denoted by $\{t_0\}$, $\{t\!_f\}$, and $\{t_n\}$, respectively.

\begin{figure}[t]
\includegraphics[angle=-90,width=\columnwidth]{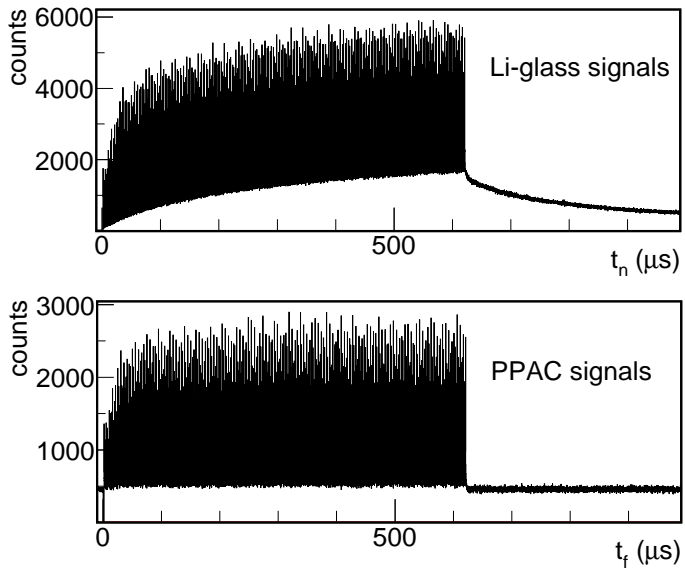}
\caption{ \label{figMacroData}
Distribution of singles events within the macro pulses for one
lithium-glass detector and one {\sc PPAC} foil.  Cuts to reduce
backgrounds have been applied (see text).}
\end{figure}
Backgrounds from several sources are present in the data.  Some of
the more significant sources are: (1) fast neutrons generated from
interactions other than fission from a particular sample foil; (2) a
"sea" of thermal and epithermal neutrons (see the rising and decaying
baseline in the top panel of fig.~\ref{figMacroData}); (3) sensitivity of
the neutron detectors to gamma rays; and (4) a small residual sensitivity
of the {\sc PPAC} to the large alpha decay rate of the plutonium samples
(fig.~\ref{figMacroData}, lower panel).

These backgrounds remain significant, even after applying some simple
background reduction cuts, such as a cuts to separate neutrons from gamma
rays in the lithium-glass detectors (exploiting the +4.78~MeV Q-value
of the $^6$Li($n,t)\alpha$ reaction), and also {\sc PPAC} pulse-height
cuts to optimize fission--alpha-decay separation.  We anticipate using
a more complete set of cuts for a final analysis of the data from these
experiments, but we still expect many components of the background
to remain.

\section{Coincidence Analysis}

Given the lists of time stamps, $\{t_0\}$, $\{t\!_f\}$, $\{t_n\}$ for the
singles data from each detector, two methods of coincidence analysis are
presented to fully identify the triplet of times $\{t_0,t\!_f\!,t_n\}$,
of a fission event.  Both methods identify two lists of time doublets
(or coincidences).  The triplet of a fission event is then a double
coincidence, formed with one coincidence from each list, matched on the
same time of their common signal.  The two methods are distinguished by
whether the coincidence pair $\{t\!_f,t_n\}$ is one of the initial lists
of coincidences (method one), or implied later (method two), and which
signal is used as the common signal between the lists ($t\!_f$ in method
one, and $t_0$ in method two).  Details of each method are given below.
While method one follows the natural time ordering of detected signals
for a fission event, the discussion below shows that method two leads to a
more detailed understanding of the backgrounds due to chance coincidences.

\begin{figure}[t]
\includegraphics[angle=-90,width=\columnwidth]{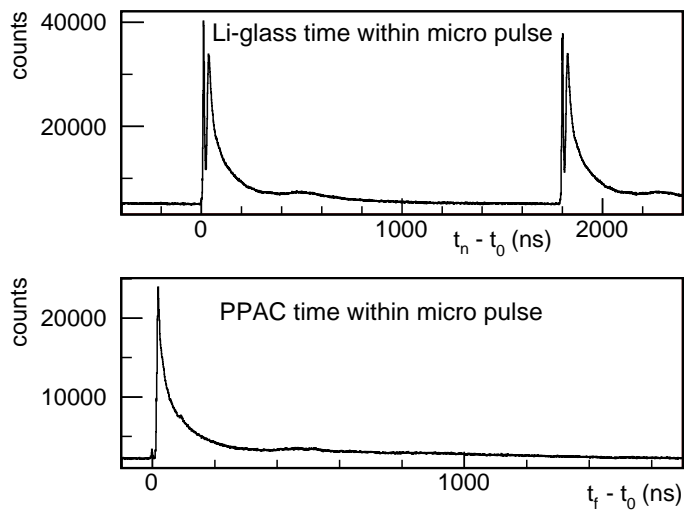}
\caption{ \label{figDoubleSingles}
Lithium glass detector (upper panel), and {\sc PPAC} (lower panel)
signal times within the micro-pulse structure.}
\end{figure}
\begin{figure}[t]
\includegraphics[angle=-90,width=\columnwidth]{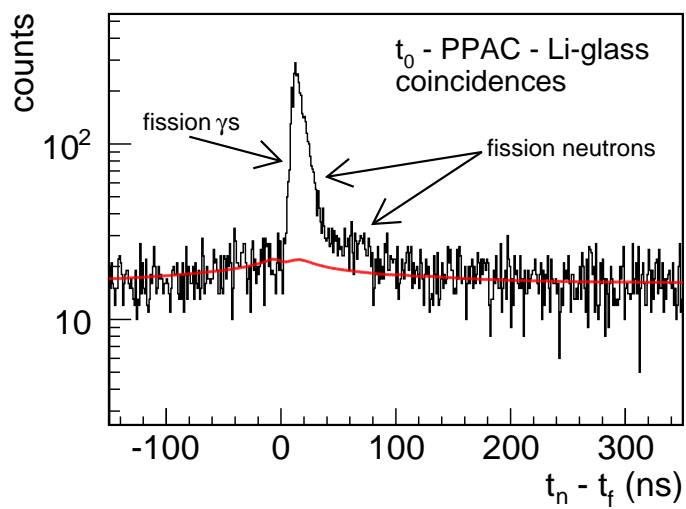}
\caption{ \label{figNftof}
Time of flight histogram (upper curve, black online) for coincidences from
one lithium glass--PPAC combination, correlated on a common $t_0$, and the
measured background from chance coincidences (lower curve, red online).}
\end{figure}
In the first method, the lists of $\{t\!_f\}$ and $\{t_n\}$, from
which fig.~\ref{figMacroData} was made, are searched to form a list
of coincidences $\{t\!_f\!, t_n\}$.  At this point a histogram of
$t_n\!-t\!_f$ could be filled, and a measurement of the background due
to chance coincidences could be derived using Eq.~\eqref{cCounts}.
The quantities $a$ and $b$ would be taken from the singles histograms
of fig.~\ref{figMacroData}, and $N$ would be the number of macro pulses
used to acquire the data.

To complete the first method, pairs of $\{t_0,t\!_f\}$ are also
identified, and then correlated with the $\{t\!_f\!,t_n\}$ pairs on
a common $t\!_f\!$.  Note that in this method, the background due
to chance-coincidences is determined before the fission event time is
correlated with the neutron production time.  This makes it difficult to
use the first method to study the background dependence on $t\!_f - t_0$.
Therefore a second method was found, which although leading less directly
to the $\{t\!_f,f_n\}$ coincidences, allows to measure the background
even with complicated cuts.

\begin{figure*}[t]
\includegraphics[angle=-90,width=\textwidth]{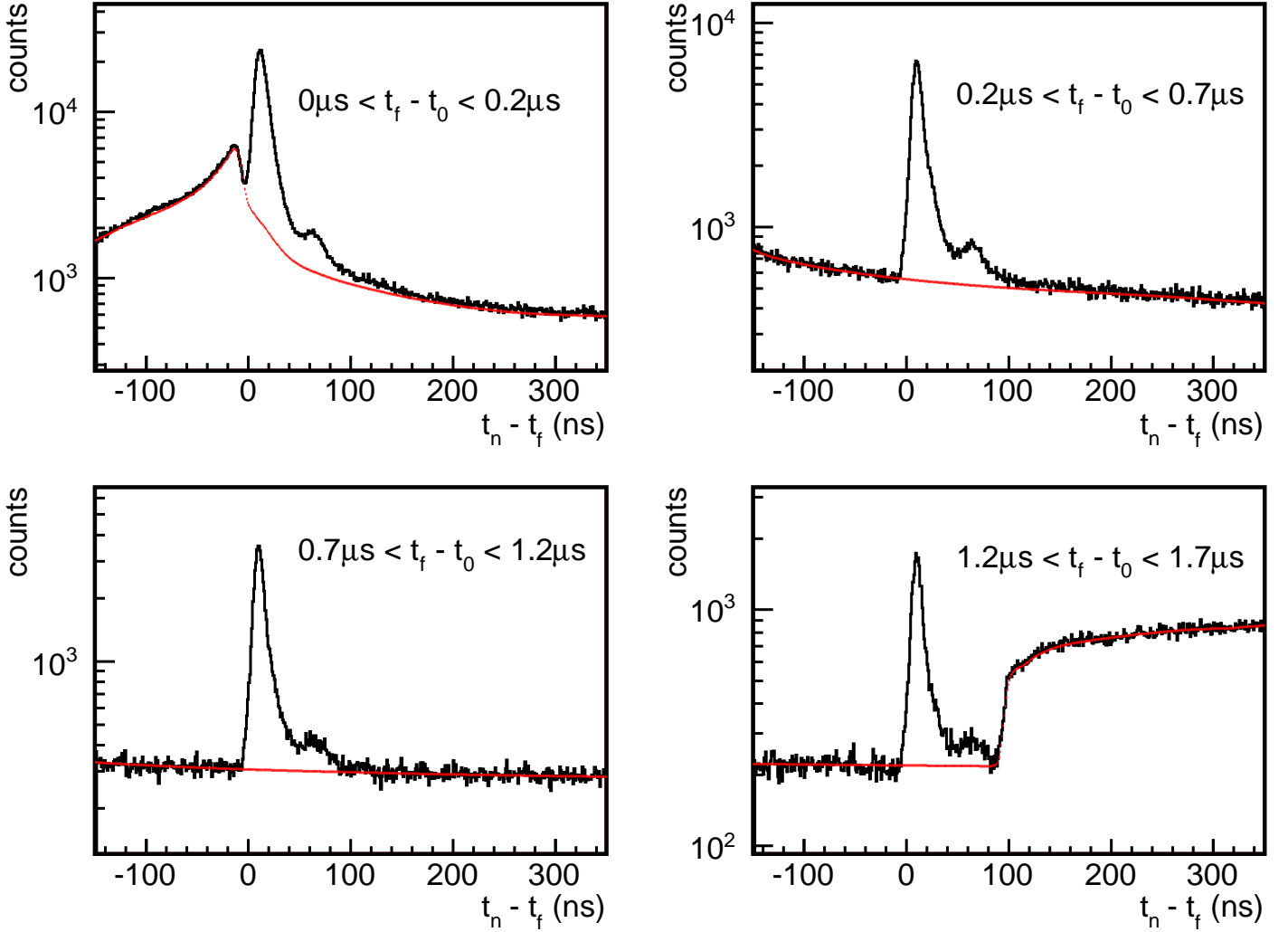}
\caption{ \label{figFourNftof}
Time of flight histogram (upper curves, black online) for coincidences
between many combinations of lithium glass--PPAC detectors correlated on
a common $t_0$, and the measured background due to chance coincidences
(lower curves, red online), for four different ranges of $t\!_f-t_0$,
and summed over all available lithium-glass--{\sc PPAC} combinations.}
\end{figure*}
For the second method, lists of $\{t_0,t\!_f\}$ between the
fission-fragment and $t_0$ signals (the same list as used in the final
stages of the first method), and between the neutron detector and $t_0$
signals $\{t_0,t_n\}$ are formed, and appropriate histograms filled
(fig.~\ref{figDoubleSingles}).  The $t_0$ times are not generated by
a Poisson process and so Eq.~\eqref{cCounts} is not yet applicable.
A second search is then used to correlate pairs of $\{t_0,t\!_f\}$
and $\{t_0,t_n\}$ on a common $t_0$, thus allowing to fill the
$t_n\!-t\!_f$ histogram (fig.~\ref{figNftof}).  $\{t_0,t\!_f\}$ and
$\{t_0,t_n\}$ are each driven by a (time-dependent) Poisson process,
and so Eq.~\eqref{cCounts} can be used to determine a background due to
chance coincidences for the histogram of $t_n\!-t\!_f$ for a common $t_0$
(red curve on fig~\ref{figNftof}).  The $a$ and $b$ values are taken
from the histograms of fig.~\ref{figDoubleSingles}, and $N$ is now the
total number of $t_0$ signals.  In contrast to the first method, the
background can still be determined even with cuts applied to the fission
time, $t\!_f-t_0$, by applying cuts to the list of $\{t_0,t\!_f\}$
before performing the second stage of searching, and then filling
the histograms of figs.~\ref{figDoubleSingles} and \ref{figNftof}.
Fig.~\ref{figFourNftof} shows such an analysis.

For the analyses both with and without cuts (figs.~\ref{figNftof}
and~\ref{figFourNftof}), attention is brought to the smaller peak in
the double coincidence data for $50{\rm ns} < t_n\! - t\!_f < 100{\rm
ns}$, the tails of which are of interest (especially the slower times,
i.e.\ $<240$~keV outgoing neutron energy)\cite{chinuPaper}.  The large
background under the low energy tail (larger $t_n\!-t\!_f$) makes this
region especially suitable for the new background measurement techniques
(see the figure-of-merit discussion below).  The larger peak ($0{\rm
ns} < t_n\! - t\!_f < 50{\rm ns}$) is dominated by fission photons,
which are to be removed by cuts in a more complete analysis, but should
also contain events from fission with higher energy outgoing neutrons.
Features of the background shapes are better understood once the full
two-dimensional background is presented below, but it is important to
note from fig.~\ref{figFourNftof}, firstly that the backgrounds are not
flat and secondly that the background shape varies with the incident
neutron energy.

As a practical matter, data processing for the background analysis may be
performed either before or after adding independent subsets of the data
set, depending on whether significant systematic changes in the count
rates have occurred or not.  Systematic changes between subsets require
independent background analyses, whereas summing over independent
subsets requires less resources (e.g.\ {\sc CPU} time and memory) to
process the data.  Both approaches work (see \ref{appSum}), or even a
combination, depending on the details of the experiment.  The current
data were acquired in four hour subsets spanning about one hundred hours
and one background analysis was performed on the summed results.

\begin{figure*}[t]
\includegraphics[angle=-90,width=\textwidth]{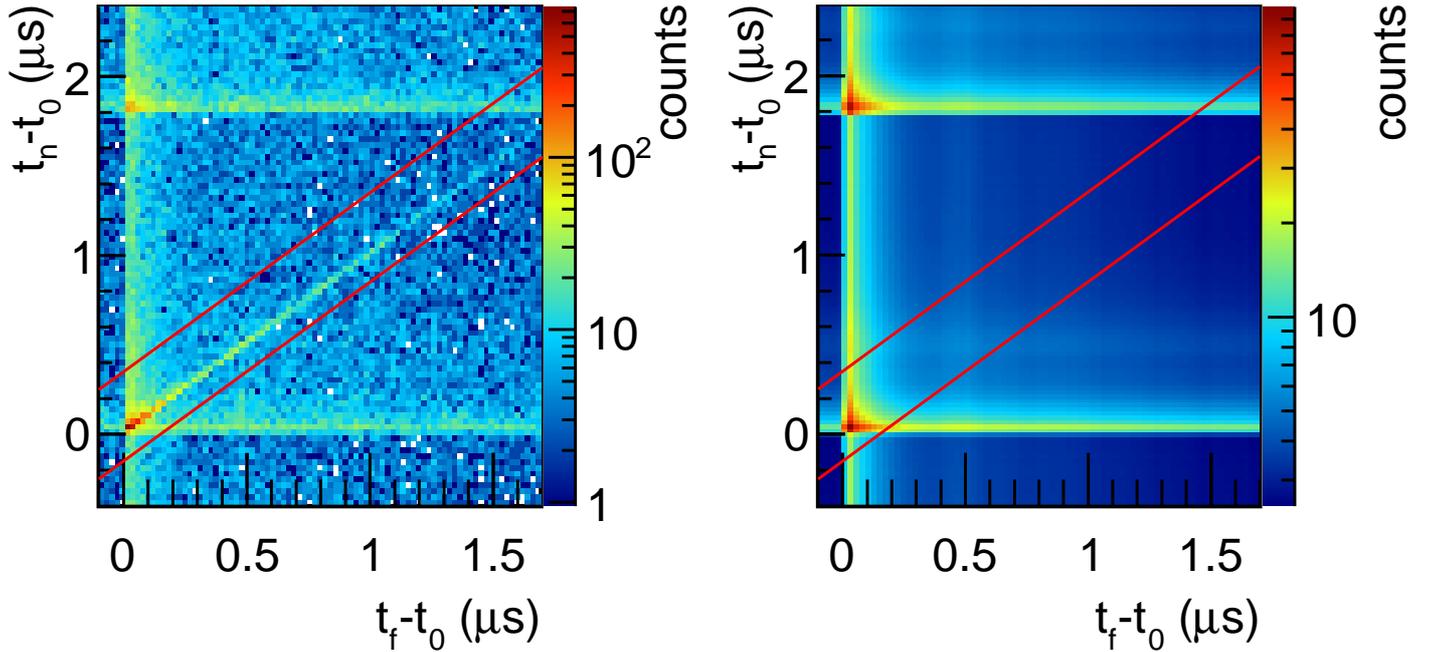}
\caption{ \label{fig2dAnalysis}
Measured two-dimensional correlations (white indicates zero counts)
between $\{t\!_f-t_0\}$ and $\{t_n\!-t_0\}$ (left), and the measured
background quantity $\gamma(t_a,t_b)$ (right).  Overlaid diagonal lines
(red online) mark the limits used to project one dimensional distributions
of fig.~\ref{figNftof} (and~\ref{figFourNftof}).}
\end{figure*}
A second practical consideration is how to handle the two-dimensional
(2D) nature of Eq.~\eqref{cCounts}.  The values of $a$ and $b$ each
depend on one or the other of the two times for the random processes,
in this case, $t\!_f-t_0$ and $t_n\!-t_0$, so that $\gamma(t_a,t_b)$
is a dyadic function of these times.  The one-dimensional (1D)
background presented in fig.~\ref{figNftof} is a projection along
lines parallel to the primary diagonal of $\gamma(t_a,t_b)$ (see
right hand side of fig.~\ref{fig2dAnalysis}), just as the coincidence
spectrum itself is a projection along the same lines of a correlation
plot between $\{t\!_f-t_0\}$ and $\{t_n\!-t_0\}$ (left hand side of
fig.~\ref{fig2dAnalysis}).  Processing the complete 2D background space
is not always necessary.  Firstly, it is only necessary to process that
section of the space near the coincidence diagonal (between the red lines
in fig.~\ref{fig2dAnalysis}).  Secondly, for most analyses a direct
determination of the 1D distribution can proceed using the formulas
of \ref{appOneBin}.  If a 2D analysis is still required (e.g. the next
four paragraphs), a lower resolution will often suffice.

Understanding the 2D nature of the problem can provide insights
into the results obtained, as features of the 1D background shape
can be attributed to summing over different features in the 2D plot
of fig.~\ref{fig2dAnalysis}.  These features will become more or less
dominant depending on the summation regions used due to cuts (such as
those on $t\!_f-t_0$ in fig.~\ref{figFourNftof}), and also due to the
sensitivity of each detector to the physical processes contributing to
the background.

For example, one feature of the 1D background in fig.~\ref{figNftof}
is a modestly peaked structure at small time differences.  The 2D
distribution shows that the small 1D peak arises from a very high
and narrow peak, located at the intersection of the high count rate
bands near $t\!_f-t_0 \approx 0$, and $t_n\! - t_0 \approx 0$, with tails
following along each of the band.  The peak is then a result of chance
coincidences between fission neutrons (or other beam induced neutrons)
and fission signals.  The upper left panel of fig.~\ref{figFourNftof}
shows a much stronger peak, now shifted to slightly negative $t_n\!-t\!_f$.
In this case, the two most forward angle neutron detectors (not included
in fig.~\ref{fig2dAnalysis}) are very sensitive to direct scattering of
neutrons from the beam and would show up in the two-dimensional analysis
as a strong enhancement of the horizontal band near $t_n\!-t_0\approx
0\mu{\rm s}$, and the 1D peak arises from chance coincidences between
the scattered beam neutrons and fissions\cite{scatterBackground}.
Similarly, the step in the background in the lower right panel of
fig.~\ref{figFourNftof} arises from the scattering into these two
detectors, now from beam neutrons from the next micro pulse at $t_n\!-t_0
\approx 1.8\mu{\rm s}$.

On the other hand, the low background in the center of the 2D distribution
sums over many channels in the 1D projection, and can therefore become
comparable in magnitude to the peak at $t_n\!-t_0 \approx t\!_f-t_0 \approx
0$.  The instantaneous neutron beam intensity is reduced in this region.
Therefore chance coincidences in this region are more likely to involve
neutrons from the "sea" of thermal and epithermal neutrons, and/or the
plutonium alpha decays events in the {\sc PPAC}.

Another striking aspect of the background measurement in
figs.~\ref{figNftof} and~\ref{figFourNftof} is how small the statistical
fluctuations on the background measurements are.  In fact, the one-sigma
uncertainty bars are marked on the figure, and even on the log plot
are hard to see.  The origin of the small uncertainties derives from
both the large number of counts available in the two input spectra,
and also from dividing by a large value of $N$ (in this case $N\approx
2.5 \times 10^9$).  A full statistical analysis, including a derivation
of the covariance matrix between elements of the background measurement
is presented in \ref{appCorr}.

The method is extensible to more complicated analysis, such as calculating
the background as a function of a singles parameter against the time
difference of the coincidences; or even then summing over the time
difference coordinate.  For example, full analysis of the example
experiment will apply a kinematic cut in the outgoing time-of-flight
versus outgoing neutron pulse height space to remove the photon peak
from figs.~\ref{figNftof} and~\ref{figFourNftof}, requiring summing the
background contribution only in the same region.  Other possibilities,
perhaps more suitable for other experiments, include projecting the
background onto the axis of the singles parameter (or combination of
singles parameters); or to extend Eq.~\ref{cCounts} to higher multiplicity
coincidences.

\section{Figure of Merit}

We now consider the impact of these results when deciding how much time
to spend acquiring data, by contrasting the signal-to-noise ratio of
traditional methods with that for the new method.

With a system that does not record all the singles data, it is often
necessary to dedicate some fraction, $\tau$, of the time to measuring
the foreground (with background), and the remaining fraction ($1-\tau$)
to measuring the background alone.  If the strength of the background
relative to the foreground is $\phi$, then the number of counts, $c_{f+b}$
and $c_b$ obtained during the two parts of the measurement are
\addParIfLineNumbers
\begin{align}
c_{f+b} &= \cOne(1+\phi)\tau \pm \sqrt{\cOne(1+\phi)\tau} ,
\\
c_b     &= \cOne\phi(1-\tau) \pm \sqrt{\cOne\phi(1-\tau)} ,
\end{align}
where $\cOne$ is a characteristic constant of the experiment (representing
the number of foreground counts if no time is dedicated to the background
measurement, i.e.~$\tau=1$, any $\phi$).  The number of foreground
counts, $c_f$, is obtained from these equations by normalizing $c_b$
and subtracting from $c_{f+b}$:
\begin{equation}
\begin{split}
c_f &= c_{f+b} - \frac{\tau}{1-\tau}c_b
\\
    &= \cOne\tau \pm \sqrt{\frac{\cOne\tau(1+\phi-\tau)}{1-\tau}}.
\end{split}
\end{equation}
The signal-to-noise ratio, $S_t = c_f / \sigma_{c_f}$, for the measurement
follows:
\begin{equation}\label{stRatio}
S_t = \sqrt{\frac{\cOne\tau(1-\tau)}{1+\phi-\tau}}.
\end{equation}

The optimal value of $\tau$ for such an experiment is obtained when the
partial derivative of the signal-to-noise ratio, with respect to $\tau$,
is zero:
\begin{equation}\label{tauOpt}
\tau = 1 + \phi - \sqrt{\phi(1+\phi)}.
\end{equation}

The small and large background limits of Eq.~\eqref{tauOpt} may be
anticipated intuitively.  Firstly, if there is no background, all the time
should be used to measure the foreground, i.e.\ $\phi=0$ implies $\tau=1$.
Secondly, at the other extreme, $\lim_{\phi\rightarrow\infty}(\tau) = 1/2$
implies to never spend more than half the time measuring the background.
Further implications of eqs.~\eqref{stRatio} and~\eqref{tauOpt} are
discussed in \cite{knoll}.

We now contrast this with an experiment designed to acquire all the
singles data from which the coincidences were identified.  The maximum
amount of time is dedicated to measuring the foreground, which gives
increased foreground statistics; the background is truly measured in situ,
which decreases systematic uncertainties in the background measurement;
and the background is measured from singles data, which has relatively
small statistical uncertainty.

If the singles rates are large, then the product $ab$ of
Eq.~\eqref{cCounts} has a relatively small uncertainty, and if $N$ is also
large, then the statistical uncertainty on $\gamma$ can be insignificant
compared to the statistical uncertainty, $\sqrt{\cOne(1+\phi)}$, of
the measurement.  The signal-to-noise ratio of a complete measurement
analyzed according to Eq.~\eqref{cCounts} is then
\begin{equation}\label{snRatio}
S_n = \sqrt{\frac{\cOne}{1+\phi}}.
\end{equation}

\begin{figure}[t]
\includegraphics[angle=-90,width=\columnwidth]{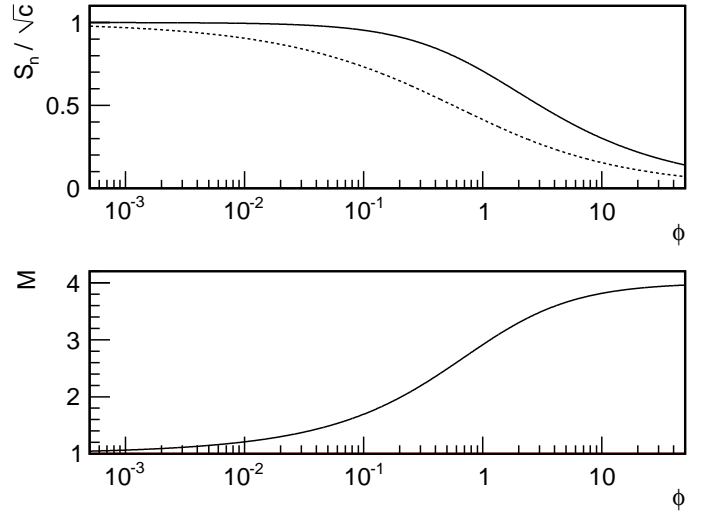}
\caption{ \label{figMerit}
Upper plot~--- signal-to-noise ratio as a function of the background
strength, $\phi$, from the complete singles method (solid line), and
traditional methods (dotted line).  Lower plot~--- figure of merit, $M$,
from Eq.~\eqref{merit}.}
\end{figure}

A useful figure of merit, $M\!$, for this new method is how much longer
an incomplete singles data set and separate background measurement
would take, to obtain the same uncertainty as a complete measurement
with background derived from singles data, i.e.
\begin{equation}\label{merit}
M = \left(\frac{S_n}{S_t}\right)^2 = \frac{1+\phi-\tau}{(1+\phi)(1-\tau)\tau}.
\end{equation}
$M$ is greater than one for any appreciable background; and for large
$\phi$, where $\tau=1/2$, $M$ can easily approach four as $M_{\tau=1/2}
= (2+4\phi)/(1+\phi)$.  The figure of merit from Eq.~\eqref{merit},
using the optimal $\tau$ from Eq.~\eqref{tauOpt}, is graphed in
fig.~\ref{figMerit}.

\section{Validity of Chance Coincidence Equations}

The effective live-time of a {\sc DAQ} with significant readout or
conversion times can be reduced by selectively triggering the system
on a subset of events~--- perhaps even triggering on only the desired
coincidence events.  It is very difficult to apply Eq.~\eqref{cCounts}
to such data sets.  Suitably complete singles data sets should not include
biases from the embedded coincidences, and ideally will not need any
significant dead-time corrections.  Therefore suitable data sets must
be acquired using a non-selective trigger~--- in direct contradiction
to a coincidence {\sc DAQ} system with a large dead time.

\begin{figure}[t]
\includegraphics[angle=-90,width=\columnwidth]{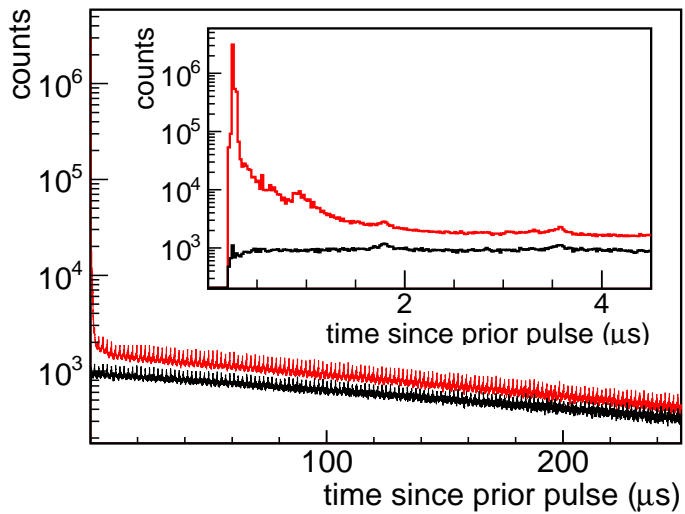}
\caption{ \label{figCleanLigl}
Time difference between adjacent peaks in one lithium-glass detector.
Upper curves (red online) show raw data, and lower curves (black online)
show data after extra peaks are removed (see text).}
\end{figure}
A practical system to acquire a complete data set requires detectors and
analog electronics with a relatively fast decay time to avoid pile up;
and a dead-timeless {\sc DAQ}.  Flash analog-to-digital converters, with
buffered multi-sample readout are ideally suited to this technique.  Self
triggered, on-board processing capabilities to identify and characterize
peaks are not strictly required, but for many experiments will be the
only practical way to obtain sufficient coincidence statistics.

It is also necessary to ensure that a data set is not over-complete.
Fig.~\ref{figCleanLigl} shows the distribution of the time differences
between each peak and the previous for one of the lithium-glass detectors,
over two different time scales.  The firmware was configured to search for
200ns wide peaks, which is then the dead time, $t_d$.  The decay constant
of the time between adjacent cleaned peaks gives a count rate in the range
of 5ks$^{-1}$.  The product of the rate and the dead time is then about
0.001, which is sufficiently less than one that eqs.~\eqref{rcCounts}
and~(\ref{cCounts}) are valid.   Peaks at 1.8, 3.6, 5.4$\mu$s, ... arise
from the real change in the count rate during micro pulses.  The raw data
curves (upper curves, red online) show other structures and extraneous
data for time differences less than 1.5$\mu$s.  These arise predominantly
from the peak searching algorithm re-triggering on fluctuations in the tail
of large peaks, and must be removed by a suitable set of cuts on the data
before applying Eq.~\eqref{cCounts} as in the lower curves (black online).
The cuts should be applied before searching for coincidences, so that
the background and signal are determined under the same conditions.

\section{Conclusions}

A method for measuring the background due to random coincidences in
coincidence experiments has been presented.  Although the method is
based on a well known formula, we describe it as a new method, as it
requires the use of a {\sc DAQ} which can record all the singles
data from which the coincidences are derived.  The ability to take such
data has only recently become practical as we now start to use waveform
digitizers, which run continuously during time periods of interest.
The dead times are now driven by detector response functions, rather
than the digitization process or data recording process.

The method is very powerful, as it allows for true in situ background
measurements, with a high degree of statistical certainty.  The term
\begin{it}in situ\end{it} applies to the physical environment in which
the detectors are located, and also to the application of the same data
analysis cuts used to search for coincidences.  The high statistical
accuracy, even for experiments with a very small background, follows
directly from an analysis procedure starting from sets of complete singles
data from which the coincidences have been identified.  This contrasts
with traditional background measurement techniques, in which a separate
set of coincidence measurements are made, with a change to the physical
environment of the experiment, such that the possibility of forming a
desirable coincidence has been removed.

A figure of merit for the method has been derived, based just on the
statistical improvement compared to traditional methods of measuring
backgrounds.  Factors of up to four times more effective use of time
are realistic.  Impressive as this figure of merit is, it is still
an underestimate of the power of the current technique.  The current
technique removes the systematic uncertainty in normalizing the background
measurement, as the conditions are now known to be truly the same as
during the foreground measurement.  This systematic improvement in the
background measurement has not been included in the figure of merit.

Furthermore, the algorithms used to extract the background are a form
of analytic data reduction.  They make no prior assumptions about the
shape of the background, and do not involve any free parameters (in
particular they do not use any type of fitting).  As such, they are
capable of reducing the data to determine very complicated background
shapes, even while maintaining large figures of merit.

It is anticipated that the method will be of use to a large class of
coincidence experiments, in which backgrounds cannot be fully eliminated.
One such set of experiments is the measurement of the prompt fission
neutron spectrum being performed at {\sc WNR/LANSCE}.  The method
has been demonstrated using preliminary data sets from this project.

\section*{Acknowledgments}

The author thanks R.~C.\ Haight for informative discussions, and all
personnel working on the Chi-Nu project\cite{chinuPaper} for allowing
access to preliminary data.  This work benefited from the use of
the {\sc LANSCE} accelerator facility and was performed under the
auspices of the US Department of Energy by Los Alamos National Security,
LLC under contract DE-AC52-06NA25396.
% and by Lawrence Livermore National Security, LLC under contract DE-AC52-07NA27344.

\appendix
\section{Summing Over Similar Data Sets}
\label{appSum}

As a practical matter, we must consider the application of
Eq.~\eqref{cCounts} under the scenario of summing two data sets, taken
sequentially, but otherwise under the same conditions.  We use subscripts
$1$ and $2$ to identify the two data sets.

Adding the background histograms directly gives
\addParIfLineNumbers
\def\csSumOneMath{
\label{csSumOne}
\gamma_{1+2} =
\left(\frac{a_1b_1}{N_1} + \frac{a_2b_2}{N_2}\right)
\\
\pm \sqrt{\frac{a_1b_1(a_1 + b_1)}{N_1^2} + \frac{a_2b_2(a_2 + b_2)}{N_2^2}},
}
\ifthenelse{\jtype=5}{ \begin{multline} \csSumOneMath \end{multline}}{ \begin{equation} \csSumOneMath \end{equation}}
whereas adding the singles histograms, and then calculating the background
gives
\addParIfLineNumbers
\def\csSumTwoMath{
\label{csSumTwo}
\gamma_{1+2} =
\frac{(a_1+a_2)(b_1+b_2)}{N_1 + N_2}
\\
\pm \sqrt{ \frac{(a_1 + a_2) (b_1 + b_2) (a_1 + a_2 + b_1 + b_2) }{ (N_1 + N_2)^2} }.
}
\ifthenelse{\jtype=5}{ \begin{multline} \csSumTwoMath \end{multline}}{ \begin{equation} \csSumTwoMath \end{equation}}

It is not immediately clear that eqs.~\eqref{csSumOne}
and~\eqref{csSumTwo} are the same result~--- indeed, at first glance
they look very different.  This dilemma is resolved by recalling that the
number of counts in a data set is proportional to the time spent acquiring
the data.  Defining the detector dependent constant of proportionality
as $k_a$ or $k_b$ so that $a = k_a N_j$ and $b = k_b N_j$, we find that
eqs.~\eqref{csSumOne} and~\eqref{csSumTwo} both reduce to the same result,
\begin{equation}\label{csSumEither}
\gamma_{1+2} = k_ak_b(N_1+N_2) \pm \sqrt{k_a k_b (k_a + k_b) (N_1 + N_2)},
\end{equation}
as desired.

\section{One bin of the projected 1D background}
\label{appOneBin}

To determine a {\sc 1D} time-difference histogram for the random
coincidence background, it helps to make explicit the time
dependence of the quantities in Eq.~\eqref{cCounts},
\begin{equation}\label{cCountsTime}
\gamma(t_a,t_b) =
 \frac{a(t_a)b(t_b)}{N} 
 \pm \frac{1}{N}\sqrt{a(t_a)b(t_b)\big(a(t_a)+b(t_b)\big)}.
\end{equation}

A single bin of $\beta(\delta_t)$, is then a sum over independent
elements of $\gamma(t_a,t_b)$ along the diagonal line $t_b=t_a+\delta_t$
(upper diagonal line, dark blue online, in fig.~\ref{figGammaCor}).
The fluctuations from individual $\gamma(t_a,t_b)$ add in quadrature,
so that
\addParIfLineNumbers
\begin{equation}\label{bOneBinCC} 
\begin{split}
\beta(\delta_t) &= \sum_{t_a} \gamma(t_a,t_a+\delta_t)
\\
  &= \frac{1}{N} \sum_{t_a} a(t_a)b(t_a+\delta_t)
\ifthenelse{\jtype=5}{ \\ &\qquad}{}
  \pm \frac{1}{N} \sqrt{\sum\nolimits_{t_a} a(t_a)b(t_a+\delta_t)\big(a(t_a)+b(t_a+\delta_t)\big)}.
\end{split}
\end{equation}

\section{Correlations in the Projected 1D Background}
\label{appCorr}

\begin{figure}[t]
\newlength{\tmplen}\setlength{\tmplen}{0.5\columnwidth}
\hfil\includegraphics[angle=-90,width=\tmplen]{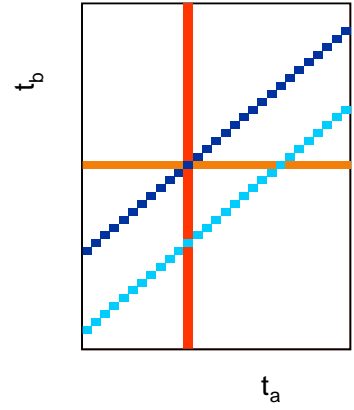}
\caption{
\label{figGammaCor}
Properties of $\gamma$ and $\beta$.  $\beta(\delta_{t1})$ and
$\beta(\delta_{t2})$ are sums of $\gamma$ along the upper and lower
diagonal lines respectively (dark and light blue online).  Correlations
arise along vertical lines (red online) of constant $a(t_a)$ and
horizontal lines (orange online) of constant $b(t_b)$.}
\end{figure}
The background in a second bin of $\beta$, sums along a different
set of $\gamma$ values, such as the lower diagonal line (light blue
online) in fig~\ref{figGammaCor}.  One term of $\beta(\delta_{t1})$
has the same value of $a(t_a)$ as one term of $\beta(\delta_{t2})$.
These terms lie on the intersection of the diagonal lines (blue online)
in fig.~\ref{figGammaCor} with the vertical line (red online) of constant
$a(t_a)$.  There is a similar correlation, with a different term of
$\beta(\delta_{t1})$, now along the horizontal line (orange online),
for a constant $b(t_b)$.  In general, these two correlations occur for
each term of $\beta(\delta_t)$, and lead to a full a covariance matrix,
${\rm Cov}\beta(\delta_{t1},\delta_{t2})$.

Expressions for ${\rm Cov}\big(\beta(\delta_{t1}),\beta(\delta_{t2})\big)$
can be derived using the methods commonly described in text books (see
e.g.~\cite{cowan}).  However such an approach is overly cumbersome,
as simplifications are easily found, reflecting the simple geometries
of fig.~\ref{figGammaCor}.

First we observe that in Eq.~\eqref{bOneBinCC}, the partial derivatives
$\partial\beta(\delta_t) / \partial\gamma(t_a, t_a+\delta_t)$ are one
(or zero), reflecting that the vertical (red online) and horizontal
(orange online) lines intersect the upper diagonal (dark blue online)
line of fig.~\ref{figGammaCor} at only one point.
\addParIfLineNumbers
\def\covbcovgMath{
\label{covbcovg}
{\rm Cov}\big(\beta(\delta_{t_1}),\beta(\delta_{t_2})\big) =
\\
\sum_{t_{a_1}} \sum_{t_{a_2}}
{\rm Cov}\big(\gamma(t_{a_1}, t_{a_1}+\delta_{t1}),
\gamma(t_{a_2}, t_{a_2}+\delta_{t2})\big).
}
\ifthenelse{\jtype=5}{ \begin{multline} \covbcovgMath \end{multline}}{ \begin{equation} \covbcovgMath \end{equation}}
Next we observe that frequently ${\rm Cov}\big(\gamma(t_{a_1},
t_{a_1}+\delta_1), \gamma(t_{a_2}, t_{a_2}+\delta_2)\big)$ is zero,
reflecting that the vertical (red online) and horizontal (orange online)
lines of fig.~\ref{figGammaCor} intersect the lower diagonal line (light
blue online) at just two points.  With these zeroes, the double summation
is reduced to just one or two single summations:
\addParIfLineNumbers
\begin{multline}\label{bCovCC}
{\rm Cov}\big(\beta(\delta_{t_1}),\beta(\delta_{t_2})\big) =
\\[\the\medskipamount]
\begin{cases}
  \displaystyle\sum_{t_a=1}^{\max(t_a)}\!  {\rm Cov}\big(\gamma(t_a,t_a+\delta_{t_1}), \gamma(t_a,t_a+\delta_{t_2})\big)
  & \text{\hbox to 0pt{\hss if $\delta_{t_1}\! = \delta_{t_2}$,}}
  \\
  \\
  \displaystyle\sum_{t_a=1}^{\max(t_a)}\!  {\rm Cov}\big(\gamma(t_a,t_a+\delta_{t_1}), \gamma(t_a,t_a+\delta_{t_2})\big)
  & \text{\hbox to 0pt{\hss otherwise.}}
  \\
  \qquad +\!\!\!\!\displaystyle\sum_{t_a=\delta_{t_2}-\delta_{t_1}}^{\max(t_a)}\!\!\!\!\!\!
  {\rm Cov}\big(\gamma(t_a,t_a+\delta_{t_1}), \gamma( t_a+\delta_{t_1}\!-\delta_{t_2},t_a+\delta_{t_1})\big)
\end{cases}
\end{multline}
Finally we use
\addParIfLineNumbers
\begin{multline}\label{ccCov}
{\rm Cov}\big(\gamma(t_{a_1},t_{b_1}), \gamma(t_{a_2},t_{b_2})\big) =
\\[\the\medskipamount]
\begin{cases}
  \displaystyle{\frac{a(t_{a})b(t_{b})\big(a(t_{a}) + b(t_{b})\big)}{N^2\strut}}
  & \text{if $t_{a_1}\! = t_{a_2} \equiv t_a$, $t_{b_1}\! = t_{b_2} \equiv t_b$,}
  \\
  \displaystyle{\frac{a(t_{a_1}) a(t_{a_2}) b(t_{b})}{N^2\strut}}
  & \text{if $t_{a_1}\! \ne t_{a_2}$, $t_{b_1}\! = t_{b_2} \equiv t_b$,}
  \\
  \displaystyle{\frac{a(t_{a}) b(t_{b_1}) b(t_{b_2})}{N^2\strut}}
  & \text{if $t_{a_1}\! = t_{a_2} \equiv t_a$, $t_{b_1}\! \ne t_{b_2}$,}
  \\
  0\strut & \text{otherwise,}
\end{cases}
\end{multline}
(obtained by standard techniques from Eq.~\eqref{cCountsTime}) to get
the desired covariance matrix, in terms of the counts in the singles spectra:
\addParIfLineNumbers
\begin{multline}\label{bCovC}
{\rm Cov}\big(\beta(\delta_{t_1}),\beta(\delta_{t_2})\big) =
\\
\qquad\frac{1}{ N^2}\sum_{t_a=1}^{\max(t_a)}
a(t_a)b(t_a+\delta_t)\big(a(t_a+\delta_{t1}-\delta_{t2})+b(t_a+\delta_{t1})\big)
\end{multline}

\section*{References}

\bibliographystyle{elsarticle-num}
\bibliography{coincBackground}

\end{document}